# Ergodic and Outage Performance of Fading Broadcast Channels with 1-Bit Feedback

Bo Niu, *Student Member, IEEE,* Osvaldo Simeone, *Member, IEEE,* Oren Somekh, *Member, IEEE,* and Alexander M. Haimovich, *Senior Member, IEEE*

*Abstract*—In this paper, the ergodic sum-rate and outage probability of a downlink single-antenna channel with $K$ users are analyzed in the presence of Rayleigh flat fading, where limited channel state information (CSI) feedback is assumed. Specifically, only 1-bit feedback per fading block per user is available at the base station. We first study the ergodic sum-rate of the 1-bit feedback scheme, and consider the impact of feedback delay on the system. A closed-form expression for the achievable ergodic sum-rate is presented as a function of the fading temporal correlation coefficient. It is proved that the sum-rate scales as $\log \log K$, which is the same scaling law achieved by the optimal non-delayed full CSI feedback scheme. The sum-rate degradation due to outdated CSI is also evaluated in the asymptotic regimes of either large $K$ or low SNR. The outage performance of the 1-bit feedback scheme for both instantaneous and outdated feedback is then investigated. Expressions for the outage probabilities are derived, along with the corresponding diversity-multiplexing tradeoffs (DMT). It is shown that with instantaneous feedback, a power allocation based on the feedback bits enables to double the DMT compared to the case with short-term power constraint in which a dynamic power allocation is not allowed. But, with outdated feedback, the advantage of power allocation is lost, and the DMT reverts to that achievable with no CSI feedback. Nevertheless, for finite SNR, improvement in terms of outage probability can still be obtained.

*Index Terms*—Ergodic sum-rate, outage probability, limited feedback, feedback delay, outdated channel state information, multiuser diversity, scaling law, power allocation, diversity-multiplexing tradeoff, Rayleigh fading.

## I. INTRODUCTION

IN wireless communications, ergodic capacity and outage probability/capacity are key information-theoretical notions to assess system performance in fading channels [1]. Ergodic capacity typically applies to non-real-time data services and measures the maximum long-term achievable rate on a time-varying fading channel (obtained by encoding over multiple independent fading realizations). On the other hand, the concept of outage is generally appropriate for delay-sensitive applications, and the outage probability is defined as the probability that the transmission rate (outage capacity) is larger than the instantaneous rate supported by the channel [1]. In fading broadcast channels, both the ergodic capacity region and the outage probability/capacity region are studied in [2] and [3], respectively, when the channel state information (CSI) is known to the transmitter and all receivers. For the case of no CSI at the transmitter side, the ergodic capacity region and outage performance are analyzed in [4] and [5], respectively.

When considering multiuser scenarios over fading channels, e.g., a cellular network, where the channel between base station and each user experiences independent time variations due to fading, a further key idea is multiuser diversity. It capitalizes on channel fluctuations in order to enhance the throughput of downlink or uplink transmissions. In fact, it has been proved that in cellular systems, serving the user with the best instantaneous channel is optimal in terms of ergodic sum-rate [6] [7]. Given this optimal scheduling that leverages multiuser diversity, reference [8] shows that the ergodic sum-rate capacity of a downlink system with Rayleigh fading channels scales as $\log \log K$ with the number of users $K$, when the instantaneous CSI for all users is known at the transmitter. With regard to the outage performance, reasoning on the dual problem of the multiple access channel [9], [10] shows that in the broadcast channel, when full CSI is available at the base station, and with a fixed transmission rate, the optimal diversity-multiplexing tradeoff (DMT) [11] is $d(r) = K(1-r)^+$,[1] where $d$ and $r$ are the diversity and multiplexing gain, respectively.

There are two problems inherent in the optimal multiuser diversity based scheduling discussed above: 1) the large amount of required feedback, and 2) the *feedback delay* that may cause the CSI fed back to the base station to be outdated. As far as the first point is concerned, various schemes have been proposed in order to reduce the feedback overhead while, at the same time, seeking to preserve the ergodic capacity [12]–[17]. Among them, a common approach prescribes feedback of a quantized version of the CSI (see review in [18]). Recently, a scheme with feedback limited to 1-bit (without considering

The work of B. Niu and A. M. Haimovich was supported in part by the National Science Foundation under Grant CNS-0626611. The work of O. Simeone was supported in part by the National Science Foundation under Grant CCF-0914899. The work of O. Somekh was supported in part by the Marie Curie Outgoing International Fellowship within the 6th European Community Framework Programme. The material in this paper was presented in part at the Asilomar Conference on Signals, Systems and Computers, Pacific Grove, CA, USA, October 2006.

B. Niu was with the Department of Electrical and Computer Engineering, New Jersey Institute of Technology, Newark, NJ 07102 USA. He is now with Apple Inc., Cupertino, CA 95014 USA (e-mail: bniu@apple.com)

O. Simeone and A. M. Haimovich are with the Department of Electrical and Computer Engineering, New Jersey Institute of Technology, Newark, NJ 07102 USA (e-mail: osvaldo.simeone@njit.edu; alexander.m.haimovich@njit.edu).

O. Somekh is with the Department of Electrical Engineering, Princeton University, Princeton, NJ 08544 USA, and with the Department of Electrical Engineering, Technion-Israel Institute of Technology, Technion City, Haifa 32000 Israel (e-mail: orens@princeton.edu).

---

[1] $(x)^+$ equals $x$ for $x \geq 0$ and 0 otherwise.

feedback delay) was proposed in [19], and further analyzed in [20]. In these works, it has been proved that the 1-bit feedback scheme preserves the optimal scaling law of $\log \log K$ (see also [21] for an extension to the MIMO case), i.e., it preserves the scaling law of multiuser diversity with full CSI feedback. For the outage case, the outage performance and DMT for the opportunistic scheme with 1-bit feedback in a $K$-user channel are studied in [22]. It is found that with a fixed transmission rate, the optimal DMT is $d(r) = K(1-r)^+$, which is the same as if full CSI were available at the transmitter.

The second problem, outdated CSI, is due to time-variations of the fading channel with respect to the feedback delay, and may cause severe degradation of the system sum-rate. In [23], the bit error rate and average data rate of an $M$-ary QAM transmission scheme are derived for broadcast fading channels with outdated, full CSI feedback. In [24], the impact of outdated CSI is studied for a selective feedback scheme, where only the users with good channel conditions (i.e., above a given threshold) feed back their full CSI. A closed-form expression for the sum-rate as a function of the fading temporal correlation $\rho$ is derived therein. With regard to the outage performance case, it is proved in [22] that opportunistic schemes, with short-term power constraints (i.e., a varying power allocation is not allowed), can not achieve an improved DMT in the presence of delayed CSI feedback.

*Main contributions and relation to previous work*

In this paper, we investigate the ergodic sum-rate and outage probability of Rayleigh fading broadcast channels with *instantaneous* and with *outdated 1-bit feedback*. We first focus on the achievable ergodic sum-rate of broadcast channels with outdated 1-bit feedback per user and fading block. Results from [19] and [20], which addressed the case of instantaneous 1-bit feedback, are obtained as a special case of the current work, when the temporal channel correlation coefficient is $\rho = 1$. The scaling law with respect to the number of users $K$ is shown to be the optimal $\log \log K$ for any non-zero temporal CSI correlation. We evaluate the sum-rate degradation due to outdated CSI in the asymptotic regimes of large $K$ and low SNR, respectively. This part of the work was first presented in [25]. Other properties of the 1-bit feedback scheme with delay for low-SNR and high-SNR asymptotics are studied as well.

We then focus on the outage performance of such channels and consider both instantaneous and outdated 1-bit feedback. We evaluate the outage probability and DMT of the proposed 1-bit feedback scheme. In contrast with [10] and [22], where the results are developed based on a short-term average power constraint, in this paper, we analyze the 1-bit feedback scheme with long-term average power constraint, for both cases of instantaneous feedback and outdated feedback. The analysis follows [26]–[28] that focused on the DMT with long-term power constraint for single-user MIMO systems. We show that power allocation is instrumental in further improving the DMT for the instantaneous feedback case. For the outdated feedback case, the DMT is shown to degrade to that of no CSI feedback, while an improvement in terms of outage probability can be observed.

## II. SYSTEM MODEL

We consider a discrete-time fading broadcast channel where a single-antenna base station communicates with $K$ single-antenna users. All users' channels are assumed to be homogeneous and experience independent block Rayleigh flat fading. Accordingly, the fading processes are independent among different users, and the fading gains remain constant during one block and may vary from block to block. The signal received by user $k$ at a given time $t$ ($t = 1, \cdots, n$, where $n$ is the block size in channel uses) within block $b$, is described as

$$y_k(t,b) = h_k(b)x(t,b) + n_k(t,b), \qquad k = 1, \cdots, K \quad (1)$$

where $h_k(b) \sim \mathcal{CN}(0,1)$ is the channel fading coefficient of user $k$ within the given fading block $b$, $n_k(t,b) \sim \mathcal{CN}(0,1)$ is complex additive white Gaussian noise with unit variance, and is assumed to be statistically independent among different users and over time. We assume that only one user is selected for transmission in each block. For the ergodic model, the channel $h_k(b)$ is an ergodic process over the blocks, and coding for each user is allowed over multiple blocks (i.e., there are no delay constraints). In the considered scheme, the transmitted signal $x(t)$ is assumed to be taken from a Gaussian codebook with per-block power constraint $\frac{1}{n}\sum_{t=1}^{n}|x(t)|^2 \leq P$. For the outage model, coding for each user is limited to a single block (strict delay constraint), and we consider both short-term (per-block) power constraint $\frac{1}{n}\sum_{t=1}^{n}|x(t)|^2 \leq P$ and long-term power constraint $E_{\mathcal{H}}[\frac{1}{n}\sum_{t=1}^{n}|x(t)|^2] \leq P$, where $x(t)$ is a function of the channel state $\mathcal{H}$ through the feedback (to be explained below).

We assume that each user is aware of its own fading power level $v_k^2(t') = |h_k(t')|^2$ based on perfect channel estimation at some time $t'$, and compares it with a prescribed threshold $\alpha$. If the fading power $v_k^2(t')$ is larger than the threshold $\alpha$, the user feeds back a single bit of "1" through a reliable uplink channel for the current fading block. Otherwise, it feeds back a single bit of "0" for the current fading block. At time $t' + \tau$, the base station receives all the feedback bits, makes its scheduling decision and starts transmitting. When there is at least one user with channel gain above the threshold, the base station randomly chooses one of the users with feedback bit "1" for transmission with power $P_1$. In case no user has a channel gain above the threshold, one user is randomly chosen and transmitted with power $P_0$. Choices of powers $P_1$ and $P_0$ are specific to different scenarios. Specifically, for the ergodic case, we assume that $P_1 = P$ and $P_0 = 0$; for the outage case, powers $P_1$ and $P_0$ are optimized with respect to the given performance criterion (outage or DMT). Note that for all the above cases, powers $P_1$ and $P_0$ are selected so as to comply with the given power constraint.

During the delay $\tau$ between perfect channel estimation and scheduling decision, the state of the channel chosen for transmission is possibly subject to change. We denote as $\rho$ the temporal channel correlation coefficient between the channels at time $t'$ and $t' + \tau$. As an example, the temporal channel correlation $\rho$ can be related to the delay $\tau$ through Jakes' model [29] as $\rho = J_0(2\pi f_D |\tau|)$, where $J_0$ is the zero-order

Bessel function of the first kind, and $f_D$ is the Doppler spread. In this paper, we assume that the base station has knowledge about $\rho$, through, e.g., estimation of the Doppler spread.

### III. ERGODIC SUM-RATE WITH OUTDATED 1-BIT FEEDBACK

In this section, we derive the ergodic sum-rate of the transmission scheme described in the previous section in the presence of outdated 1-bit feedback. We then prove that the sum-rate scaling of the 1-bit feedback scheme with delay is the same as that of the optimal non-delayed full CSI feedback scheme. Different asymptotic results are also studied at both low and high SNR regimes. In order to simplify the analysis, we assume that in case no user has fed back a "1" bit, the base station keeps silent for the current block ($P_0 = 0$). Otherwise the base station randomly chooses one of the users with feedback bit "1" and transmits with power $P_1 = P$. The average power constraint still holds. Notice that, although some benefit in terms of the sum-rate can be achieved by allocating power to the slots where no user has channel gain above the threshold, we will prove that the gain is in practice negligible, with a proper selection of the threshold and large number of users. Thus allowing silent periods does not affect much the generality of the analysis.

#### A. Ergodic Sum-rate

The main goal of this subsection is to derive the achievable ergodic sum-rate of the 1-bit feedback scheme described in Sec. II in the presence of feedback delay.

**Proposition 1** *The ergodic sum-rate with outdated 1-bit feedback given power $P$, $K$ users, temporal channel correlation coefficient $\rho$, and arbitrary threshold $\alpha$, is given by*[2]

$$R(\alpha, \rho, P, K) = \left(1 - \left(1 - e^{-\alpha}\right)^K\right) \int_0^\infty \log(1 + z^2 P) 2z$$
$$\times e^{-z^2+\alpha} Q_1\left(\frac{\sqrt{2}|\rho|}{\sqrt{1-\rho^2}} z, \frac{\sqrt{2\alpha}}{\sqrt{1-\rho^2}}\right) dz, \quad (2)$$

*where $Q_1(A, B) = \int_B^\infty x \exp\left(-\frac{x^2+A^2}{2}\right) I_0(Ax) dx$ is the first-order Marcum-Q function.*

*Proof:* At any fading block, either one or no user is selected for transmission, and long codewords (spanning multiple fading blocks) are chosen from a Gaussian codebook. The ergodic sum-rate is the product of two terms [20]: (*i*) the probability that at least one user is qualified to be chosen for transmission and (*ii*) the ergodic sum-rate for the chosen users over the fading blocks selected for transmission:

$$R(\alpha, \rho, P, K) = \Pr(N > 0) E[\log(1 + v_\tau^2 P) | v^2 > \alpha], \quad (3)$$

where $N$ is the number of users with channel power gain $v_k^2(t')$ above the threshold $\alpha$ in a given block. We have dropped the subscript $k$ due to the statistical equivalence of different users, and denoted $v = v(t')$ and $v_\tau = v(t' + \tau)$ (i.e., channel envelopes at the channel estimate and scheduling decision time instants, respectively).

[2] A natural logarithmic base is used throughout the analysis.

The probability that at least one user is qualified to be chosen for transmission is

$$\begin{aligned}\Pr(N > 0) &= 1 - \Pr(N = 0) = 1 - \left(\Pr\left(v^2 < \alpha\right)\right)^K \\ &= 1 - \left(1 - e^{-\alpha}\right)^K. \quad (4)\end{aligned}$$

In order to calculate the ergodic sum-rate for the chosen user over the fading blocks, we need the probability density function (pdf) of $v_\tau$ given the condition $v^2 \geq \alpha$. We start from the cumulative distribution function (cdf) of $v_\tau$ given the condition $v^2 \geq \alpha$,

$$\begin{aligned}F_{v_\tau}(z | v^2 \geq \alpha) &= \frac{\Pr(v_\tau < z, v \geq \sqrt{\alpha})}{\Pr(v \geq \sqrt{\alpha})} \\ &= \frac{\int_0^z dv_\tau \int_{\sqrt{\alpha}}^\infty f(v_\tau, v) dv}{\int_{\sqrt{\alpha}}^\infty 2v e^{-v^2} dv}, \quad (5)\end{aligned}$$

where $f(v_\tau, v) = \frac{4 v_\tau v}{1-\rho^2} e^{-(v_\tau^2+v^2)/(1-\rho^2)} I_0\left(\frac{2|\rho| v_\tau v}{1-\rho^2}\right)$ is the joint pdf of two correlated Rayleigh random variables [30]. By taking the derivative of (5) with respect to $z$, we achieve the conditional pdf as

$$\begin{aligned}f_{v_\tau}(z | v^2 \geq \alpha) &= \frac{\int_{\sqrt{\alpha}}^\infty \frac{4zv}{1-\rho^2} e^{-(z^2+v^2)/(1-\rho^2)} I_0\left(\frac{2|\rho| zv}{1-\rho^2}\right) dv}{e^{-\alpha}} \\ &= 2z e^{-z^2+\alpha} Q_1\left(\frac{\sqrt{2}|\rho|}{\sqrt{1-\rho^2}} z, \frac{\sqrt{2\alpha}}{\sqrt{1-\rho^2}}\right). \quad (6)\end{aligned}$$

Substituting (4) and (6) into (3) concludes the proof. ∎

#### B. Scaling Law

To gain insight into the impact of delay on the achievable ergodic sum-rate of the 1-bit feedback scheme, it is convenient to derive upper and lower bounds on (2). An upper bound on the rate is directly derived by using Jensen's inequality on (3),

$$R_{\text{up}}(\alpha, \rho, P, K) = \left(1 - \left(1 - e^{-\alpha}\right)^K\right) \log\left(1 + P\left(1 + \rho^2 \alpha\right)\right). \quad (7)$$

On the other hand, a lower bound is obtained as (see Appendix A for derivation)

$$\begin{aligned}R_{\text{low}}(\alpha, \rho, P, K) &= \left(1 - \left(1 - e^{-\alpha}\right)^K\right) \log(1 + \alpha P) \\ &\quad \left\{1 + Q_1\left(\frac{|\rho|\sqrt{2\alpha}}{\sqrt{1-\rho^2}}, \frac{\sqrt{2\alpha}}{\sqrt{1-\rho^2}}\right) \right. \\ &\quad \left. - Q_1\left(\frac{\sqrt{2\alpha}}{\sqrt{1-\rho^2}}, \frac{|\rho|\sqrt{2\alpha}}{\sqrt{1-\rho^2}}\right)\right\}. \quad (8)\end{aligned}$$

Exploiting the lower bound (8), the scaling law of the ergodic sum-rate of the 1-bit feedback scheme with respect to the number of users $K$ is stated in the following.

**Proposition 2** *For any finite power $P$ and channel correlation coefficient $\rho \neq 0$, with increasing number of users $K$, the sum-rate with outdated 1-bit feedback has the same growth rate as the full CSI feedback scheme:*

$$\lim_{K \to \infty} \frac{R(\alpha_o(K), \rho, P, K)}{\log \log K} = 1, \quad (9)$$

*where $\alpha_o(K)$ is the optimal threshold that maximizes $R(\alpha, \rho, P, K)$ for a given $K$.*

*Proof:* The lower bound (8) suggests that, in order to get a multiuser diversity gain of $\Theta(\log K)$,[3] and to make the pre-log term close to 1, a "good" choice of the threshold is $\alpha_{\text{so}}(K) = \log K - \delta$, where $\delta$ is a positive constant smaller than $\log K$ (the subscript "so" is for "suboptimal"). With this choice of threshold (termed as the suboptimal threshold), we have

$$\lim_{\substack{K \to \infty \\ \alpha = \alpha_{\text{so}}(K)}} \left(1 - \left(1 - e^{-\alpha}\right)^K\right) = 1 - e^{-e^{\delta}}, \quad (10)$$

and, as proved in Appendix B,

$$\lim_{\substack{K \to \infty \\ \alpha = \alpha_{\text{so}}(K)}} 1 + Q_1\left(\frac{|\rho|\sqrt{2\alpha}}{\sqrt{1-\rho^2}}, \frac{\sqrt{2\alpha}}{\sqrt{1-\rho^2}}\right) - Q_1\left(\frac{\sqrt{2\alpha}}{\sqrt{1-\rho^2}}, \frac{|\rho|\sqrt{2\alpha}}{\sqrt{1-\rho^2}}\right) = 1. \quad (11)$$

From (8), (10) and (11), it follows that

$$\lim_{K \to \infty} \frac{R_{\text{low}}(\alpha_{\text{so}}(K), \rho, P)}{\log \log K} = 1 - e^{-e^{\delta}}. \quad (12)$$

Since $\delta$ can be chosen any arbitrary large number (after taking $K$ to infinity), the ratio in (12) goes to 1. Therefore, since a suboptimal threshold preserves the scaling law of $\log \log K$, the ergodic sum-rate of the 1-bit feedback scheme with delay guarantees the same growth rate as the full CSI feedback scheme with an optimal threshold $\alpha_o(K)$, thus concluding the proof. ∎

**Corollary 1** *For any finite power $P$ and channel correlation coefficient $\rho \neq 0$, with increasing number of users $K$, the optimal threshold $\alpha_o(K)$ that maximizes the sum-rate $R(\alpha, \rho, P, K)$ of the 1-bit feedback scheme with delay is $\Theta(\log K)$.*

It has been proved in Proposition 2 that the ergodic sum-rate has the same capacity scaling law of $\log \log K$ as the full CSI feedback scheme. It is straightforward to see from (8) that in order to achieve the scaling law, the optimal threshold $\alpha_o(K)$ must be $\Theta(\log K)$.

*Remark 1:* With the considered power allocation $P_0 = 0$ and $P_1 = P$, it was proved that the sum-rate of the 1-bit feedback scheme with delay scales as $\log \log K$, which is the same scaling law achieved by the optimal non-delayed full CSI feedback scheme. Therefore, from a capacity scaling point of view, there is nothing to gain by allocating power to the slot where no user has channel gain above the threshold (i.e., setting $P_0 > 0$). To see this, notice that, since the optimal threshold must be $\Theta(\log K)$, the probability that no user is above the threshold is $\Pr(N = 0) = (1 - e^{-\alpha_o})^K \approx 0$, with an optimal threshold and a large number of users. Thus, allocating $P_0 > 0$ does not affect the considered performance criterion.

Another interesting asymptotic result comes from the upper bound (7). With a large number of users $K$ and optimal threshold $\alpha_o(K)$, we have

$$\lim_{K \to \infty} R_{\text{up}}(\alpha_o(K), \rho, P, K) = \log\left(1 + P\left(1 + \rho^2 \alpha_o(K)\right)\right)$$
$$\approx \log P \alpha_o(K) + 2 \log |\rho|, \quad (13)$$

where the first equality follows from the proof of Proposition 1, in which it was shown that there exists a suboptimal threshold $\alpha_{\text{so}}(K)$ such that $\lim_{K \to \infty} \left(1 - \left(1 - e^{-\alpha_{\text{so}}(K)}\right)^K\right) = 1$ (so that this holds also with an optimal threshold $\alpha_o(K)$). The approximation follows from the fact that in order to obtain a multiuser diversity of $\log \log K$, the optimal threshold $\alpha_o(K)$ is $\Theta(\log K)$, and $\log(1 + x) \approx \log x$, for $x \gg 1$. The first term in (13), $\log P \alpha_o(K)$, is the optimized asymptotic rate with a large number of users for the 1-bit feedback scheme without delay ($\rho = 0$) [20]. Therefore, the second term $2 \log |\rho|$ provides a measure of the sum-rate degradation due to feedback delay in the asymptotic regime of large $K$. In Sec. III-E, it will be show via numerical results that this quantity is in fact an accurate prediction of the actual sum-rate degradation for $K \gg 1$.

### C. Low-SNR Characterization

In this subsection we study the ergodic sum-rate of broadcast channels with outdated 1-bit feedback and operating in a power-limited (or wideband) regime. This regime is characterized by low SNR and low sum-rate. In [31] it is shown that in order to characterize the sum-rate in the low SNR regime, an affine approximation of the sum-rate versus $\frac{E_b}{N_0}$ in this regime is represented as[4]

$$\mathrm{R} \cong \frac{S_0}{3\mathrm{dB}} \left( \left.\frac{E_b}{N_0}\right|_{\mathrm{dB}} - \left.\frac{E_b}{N_0}_{\min}\right|_{\mathrm{dB}} \right), \quad (14)$$

where two parameters should be considered: (*i*) the minimum signal energy-per-information bit $\frac{E_b}{N_0}_{\min}$ required for reliable communication

$$\left.\frac{E_b}{N_0}\right|_{\min} = \left.\frac{\log 2}{\frac{\partial R}{\partial P}}\right|_{P=0},$$

where $E_b$ [joules] is the transmitted energy per information bit, and $N_0$ [watts/Hz] is the noise spectral density. (*ii*) the spectral efficiency slope $S_0$, also referred to as wideband slope, as a function of $\frac{E_b}{N_0}$, at $\frac{E_b}{N_0}_{\min}$.

$$S_0 = \left.\frac{2\left(\frac{\partial R}{\partial P}\right)^2}{-\frac{\partial^2 R}{\partial P^2}}\right|_{P=0}.$$

In this subsection, we study the performance of the outdated 1-bit feedback scheme in the wideband limit by deriving closed-form expressions for both $\frac{E_b}{N_0}_{\min}$ and $S_0$. An asymptotic analysis of the two parameters with large number of users $K$ and a suboptimal threshold $\alpha_{\text{so}}$ is also presented.

Deriving from (2), and using Eq. (B.28) in [30], we have

$$\left.\frac{E_b}{N_0}\right|_{\min} = \left.\frac{\log 2}{\frac{\partial R(\alpha, \rho, P, K)}{\partial P}}\right|_{P=0}$$
$$= \frac{\log 2}{\left(1 - (1 - e^{-\alpha})^K\right)(1 + \rho^2 \alpha)}, \quad (15)$$

---

[3]Let $f(n)$ and $g(n)$ be two positive functions. We write $f(n) = \Theta(g(n))$ if there exist positive constants $c_1$, $c_2$, $n_0$ such that for all $n \geq n_0$, $0 \leq c_1 g(n) \leq f(n) \leq c_2 g(n)$.

[4]Following [31], the notation R is introduced so as to denote the sum-rate as a function of $\frac{E_b}{N_0}$. This notion is related to the sum-rate of $R$ as a function of the SNR $P$, through $\mathrm{R}(\frac{E_b}{N_0}) = R(P)$ and $P = R(P)\frac{E_b}{N_0}$.

and

$$S_0 = \left. \frac{2\left(\frac{\partial R(\alpha,\rho,P,K)}{\partial P}\right)^2}{-\frac{\partial^2 R(\alpha,\rho,P,K)}{\partial P^2}} \right|_{P=0}$$
$$= \frac{\left(1-(1-e^{-\alpha})^K\right)(1+\rho^2\alpha)^2}{1+2\alpha\rho^2-\alpha\rho^4+\frac{1}{2}\alpha^2\rho^4}. \quad (16)$$

Using the results in (15) (16), we can quantify the sum-rate degradation due to the feedback delay in the asymptotic regime of low SNR for any number of users by means of (14).

To further analyze the above results (15) (16), we consider the asymptotic scenario with large number of users $K$ and a suboptimal threshold $\alpha_{\text{so}}(K) = \log K - \delta$ (this suboptimal threshold guarantees the asymptotic optimality of the scaling law, as proved in Proposition 2). From (15) and (16), we have for $\rho \neq 0$ and $K \gg 1$

$$\frac{E_b}{N_0}_{\min} \to \frac{\log 2}{\rho^2 \log K}, \quad (17)$$

and

$$S_0 \to 2. \quad (18)$$

Substituting these results in (14), we obtain an affine approximation of the sum-rate versus $\frac{E_b}{N_0}$ in the wideband regime with large number of users

$$R\left(\alpha,\rho,K,\frac{E_b}{N_0}\right) \cong \frac{2}{3\text{dB}} \left( \left.\frac{E_b}{N_0}\right|_{\text{dB}} - 10\log_{10}\frac{\log 2}{\rho^2 \log K} \right). \quad (19)$$

It is known that $\frac{E_b}{N_0}_{\min}$ for reliable communication over a fading channel with no CSI at the transmit side is $\log 2 = -1.59$dB. With 1-bit CSI feedback, a large number of users $K$ and a temporal channel correlation coefficient $\rho \neq 0$, the denominator in (17) shows a multiuser diversity gain of $\rho^2 \log K$. With increasing $|\rho|$, this leads to a decreasing required $\frac{E_b}{N_0}_{\min}$ for reliable communication. Regarding the spectral efficiency slope $S_0$, it can be obtained from (16) that it equals 1 when the temporal channel correlation coefficient is $\rho = 0$. This result coincides with the case of perfect receiver side information but with no CSI at the transmitter described in [31]. When $\rho \neq 0$, the asymptotic spectral efficiency slope (18) equals 2 as for an AWGN channel [31].

### D. High-SNR Characterization

Multiplexing gain (high-SNR spectral efficiency slope) is commonly used to characterize the spectral efficiency in the high SNR regime [32]. It is defined as

$$r = \lim_{P \to \infty} \frac{R(P)}{\log P}. \quad (20)$$

It is difficult to provide an explicit expression of the multiplexing gain of the ergodic sum-rate of broadcast channels with outdated 1-bit feedback, through (2). Upper and lower bounds of the multiplexing gain are given through (7) and (8) as

$$r_{\text{up}} = \left(1-(1-e^{-\alpha})^K\right), \quad (21)$$

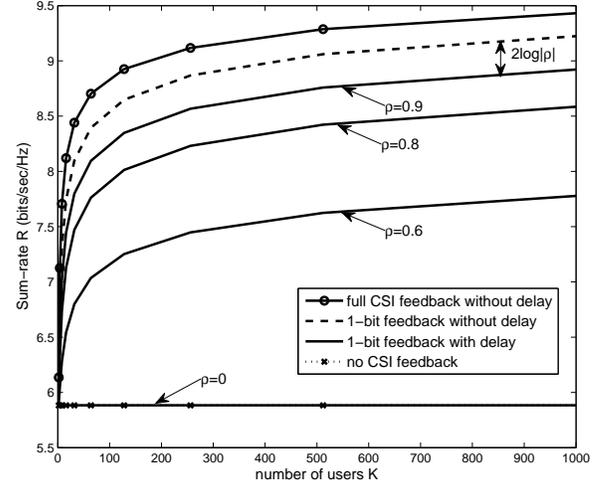

Fig. 1. Ergodic sum-rate $R$ versus the number of users $K$ with outdated 1-bit feedback for different channel temporal correlation coefficient $\rho$ and optimal threshold $\alpha_o(K)$. Ergodic sum-rate capacity with non-delayed full CSI, no CSI ($\rho = 0$), and 1-bit feedback without delay ($\rho = 1$) are also shown for reference ($P = 20$ dB).

and

$$r_{\text{low}} = \left(1-(1-e^{-\alpha})^K\right) \left\{ 1 + Q_1\left(\frac{|\rho|\sqrt{2\alpha}}{\sqrt{1-\rho^2}}, \frac{\sqrt{2\alpha}}{\sqrt{1-\rho^2}}\right) - Q_1\left(\frac{\sqrt{2\alpha}}{\sqrt{1-\rho^2}}, \frac{|\rho|\sqrt{2\alpha}}{\sqrt{1-\rho^2}}\right) \right\}. \quad (22)$$

With large number of users $K$ and the suboptimal threshold $\alpha_{\text{so}}$, both the upper bound (21) and the lower bound (22) converge to 1, which is the multiplexing gain for a SISO system.

### E. Numerical Results

Fig. 1 shows the ergodic sum-rate with outdated 1-bit feedback versus the number of users $K$ for different values of the temporal channel correlation coefficient $\rho$, with optimal threshold $\alpha_o(K)$ (evaluated numerically) and SNR $P = 20$ dB, based on the closed-form expression of the achievable ergodic sum-rate (2). The ergodic sum-rate capacity with non-delayed full CSI [8], no CSI ($\rho = 0$), and 1-bit feedback without delay ($\rho = 1$), are also shown for reference. It can be seen that in the presence of delay, the sum-rate with 1-bit feedback shows the same scaling law of the optimal transmission scheme with full CSI. Moreover, for different channel correlation coefficients, the rate degradation is well quantified by $2\log|\rho|$ as derived in Sec. III-B.

Fig. 2 shows the sum-rate with outdated 1-bit feedback as a function of $E_b/N_0$, and its wideband approximation (14) according to (15) and (16), for different temporal channel correlation coefficients $\rho$, with optimal threshold $\alpha_o(K)$ and finite number of users $K = 100$. Spectral efficiencies and their affine approximations with non-delayed full CSI, no CSI ($\rho = 0$), and 1-bit feedback without delay ($\rho = 1$) at low SNR regime, are also shown for reference. It is seen that even outdated 1-bit feedback affords a relevant multiuser diversity gain in terms of $E_b/N_{0_{\min}}$ with respect to the case of no CSI ($\rho = 0$). This gain can be quantified as $10 \log_{10}(\rho^2 \log K)$dB when the number of users $K$ is large (see (17)). The spectral efficiency slope increases from 1, which corresponds to the case of no CSI feedback, to $S_0 = 2$, which equals to the spectral efficiency slope of a Gaussian channel [31].

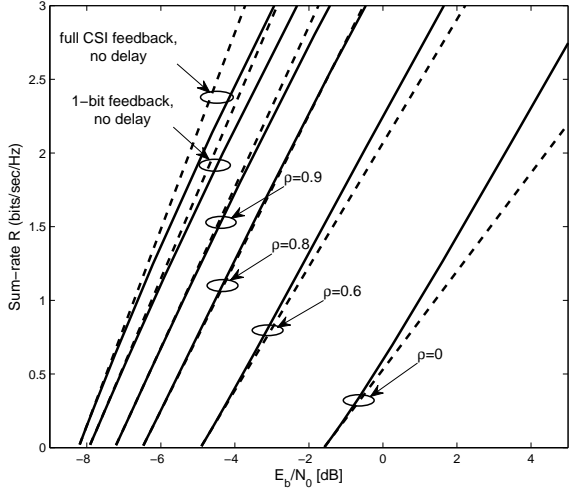

Fig. 2. Ergodic sum-rate with outdated 1-bit feedback in the low SNR regime (solid line) and its affine approximation (14) for different channel temporal correlation $\rho$ (dashed line). Sum-rate capacity with non-delayed full CSI, no CSI ($\rho = 0$), and 1-bit feedback without delay ($\rho = 1$), are also shown for reference ($K = 100$).

## IV. Outage Performance of the 1-Bit Feedback Scheme

In this section, we discuss the outage performance of the 1-bit feedback scheme. For the outage case, the channel is constant for the entire duration of the codeword. It is shown in [3] [33] [34] that power allocation has an important impact on the outage performance, compared to the case of no power allocation. In [10] and [22], it is pointed out that in the broadcast channel, with a fixed transmission rate $R$, the optimal DMT is $d(r) = K(1-r)^+$ with instantaneous full CSI or even 1-bit feedback. The results above are achieved with a short-term average power allocation assumption, and the base station transmits with the same power to the chosen user regardless the feedback bits ($P_1 = P_0 = P$). Inspired by [26]–[28], where the DMT of a point-to-point communication with limited feedback CSI and long-term power allocation is discussed, we investigate the outage performance for a 1-bit feedback scheme with long-term average power allocation in this section, and analyze how power allocation improves the DMT in the multiuser scenario [35]. By long-term average power allocation, different power $P_1$ or $P_0$ is adaptively allocated to the chosen user during each of the fading blocks, according to the CSI feedback bits from all users. Let the rate transmitted by the base station be $R$, we first consider the case with instantaneous feedback ($\rho = 1$), and derive the outage probability of the 1-bit feedback scheme with long-term average power constraint. To gain more insight into the system performance, we resort to asymptotic analysis (high SNR) of the DMT of the 1-bit feedback scheme. Then we consider the scheme with outdated CSI and long-term average power constraint, and extend the results for both outage probability and the DMT.

### A. Instantaneous 1-Bit Feedback

Recall that long-term average power constraint implies that different powers $P_1$ and $P_0$ are used for transmission, when respectively, one of the users' channel gain exceeds the threshold ($N > 0$), and when no channel gain exceeds the threshold ($N = 0$). We first analyze the outage performance of the 1-bit feedback scheme with instantaneous CSI feedback.

#### A.1. Outage Probability

The following proposition holds for the outage probability:

**Proposition 3** *The outage probability of the considered transmission scheme with instantaneous 1-bit feedback, long-term average power constraint $P$, $K$ users, threshold $\alpha \geq 0$, and a prescribed rate $R$, is given by*

$$\epsilon = \epsilon_1 \times \Pr(N > 0) + \epsilon_0 \times \Pr(N = 0), \quad (23)$$

*where*

$$\epsilon_1 = \Pr[\log(1+|h|^2 P_1) < R \,|\, |h|^2 \geq \alpha]$$
$$= \begin{cases} 0 & \text{if } R \leq \log(1+P_1\alpha) \\ 1 - \exp\left(\alpha - \frac{e^R - 1}{P_1}\right) & \text{otherwise} \end{cases}, (24)$$

*and*

$$\epsilon_0 = \Pr[\log(1+|h|^2 P_0) < R \,|\, |h|^2 < \alpha]$$
$$= \begin{cases} \frac{1 - \exp(-\frac{e^R - 1}{P_0})}{1 - e^{-\alpha}} & \text{if } R \leq \log(1+P_0\alpha) \\ 1 & \text{otherwise} \end{cases}, (25)$$

*are the probabilities of outage conditioned on the feedback bits.*

*Proof:* Since Rayleigh fading channels are considered, $|h|^2$ is an exponential distribution random variable. Hence the outage probability given that at least one user with channel gain above the threshold and feeds back a bit of "1" becomes

$$\epsilon_1 = \Pr\left(|h|^2 < \frac{e^R - 1}{P_1} \,\bigg|\, |h|^2 \geq \alpha\right) = 1 - \exp\left(\alpha - \frac{e^R - 1}{P_1}\right), \quad (26)$$

if $R > \log(1 + P_1\alpha)$; and

$$\epsilon_1 = 0,$$

if $R \leq \log(1 + P_1\alpha)$. Solving $\epsilon_0$, the outage probability given that no user has a channel above the threshold and all feed back a bit of "0", in a similar way, we have

$$\epsilon_0 = \begin{cases} \frac{1 - \exp(-\frac{e^R - 1}{P_0})}{1 - e^{-\alpha}} & \text{if } R \leq \log(1+P_0\alpha) \\ 1 & \text{otherwise} \end{cases}. \quad (27)$$

Thus the expression for outage probability can be derived using (4) and (23). ∎

#### A.2. Diversity-Multiplexing Tradeoff

The notion of DMT was introduced in [11] in order to characterize the performance of transmission schemes over block fading channels for high SNR. It essentially reveals the tradeoff between the error probability and the data rate of a system, and is represented by the multiplexing gain

$$r = \lim_{P \to \infty} \frac{R(P)}{\log P},$$

where $R(P)$ is the rate at power $P$, and the diversity gain

$$d = -\lim_{P \to \infty} \frac{\log \epsilon(P)}{\log P}, \tag{28}$$

where $\epsilon(P)$ is the outage probability at power $P$.

Previous works have shown that for a point-to-point link, instantaneous CSI feedback can dramatically improve the DMT. In particular, [26]–[28] focus on the DMT for point-to-point links with partial CSI at transmitter. As a special case of the result there, for a SISO system with instantaneous 1-bit feedback to the transmitter, the optimal DMT is $d(r) = 2(1-r)^+$. In order to achieve this result, power allocation based on the feedback is a key factor. In this subsection, we look at the DMT for broadcast channels with instantaneous 1-bit feedback per user per fading block, and show the important roles that power allocation and multiuser diversity play in determining the DMT.

**Proposition 4** *The DMT of broadcast channels with instantaneous 1-bit feedback, $K$ users and long-term average power constraint, is lower bounded by*

$$d(r) = 2K(1-r)^+. \tag{29}$$

*Proof:* We look at a specific power allocation scheme, where we transmit $P_1 = P/2$ to the chosen one user, when there is at least one user with channel gain above the threshold $\alpha$, and $P_0 = P_1/\Pr(N=0) = \frac{P}{2(1-e^{-\alpha})^K}$ to the chosen user, when no user has a channel gain above the threshold. The intuition behind this power allocation scheme is that less power is used for transmissions when the chosen user has a favorable channel. The saved power is used to lower the outage probability, when all users are under poor channel conditions. We also choose the threshold $\alpha$ to guarantee that there is no outage occur during the transmission when there is at least a user above the threshold $\alpha$. Details of the proof are found in Appendix C.

*Remark 2:* In the downlink of a multiuser scenario with $K$ users and long-term average power constraint, even with only 1-bit of CSI feedback from users to the base station, the DMT increases linearly with the number of users in the system. The key factor here is that in the multiuser scenario, the probability that all users' channel gains are below the threshold is drastically reduced compared to point-to-point communication. This explains the improvement in DMT compared to [26]–[28]. It is worth noticing that for the special case of a single user, our result reduces to the DMT of the point-to-point scenario. Also, we remark that the DMT obtained with the long-term power constraint assumed in formulating Proposition 4 is larger by a factor of two than the DMT for the case of short-term power constraint [22].

*Remark 3:* The DMT in Proposition 4, and in general the results reported in this section, are only achievable. We are currently unaware of any tight upper bound. A simple upper bound on the DMT could be found by allowing all the users fully cooperate with each other, so that the system works as a point-to-point single-input-multiple-output (SIMO) system with one transmit antenna and $K$ receive antennas. With 1-bit feedback from each receive antenna, the optimal DMT for this SIMO case can be derived based on the result from [26] as $d(r) = \sum_{m=1}^{M=2^K} K^m(1-r)^+$, which is generally larger than the achievable DMT of Prop. 4, $d(r) = 2K(1-r)^+$. We also remark that more complicated schemes can be derived to achieve the same DMT of $d(r) = 2K(1-r)^+$, but with better outage probability result, e.g., the three-level power allocation scheme proposed in [26].

### B. Outdated 1-Bit Feedback

In this subsection, we extend the outage performance analysis to the scenario with outdated 1-bit feedback and long-term average power constraint.

*B.1. Outage Probability*

**Proposition 5** *The outage probability of the considered transmission scheme with outdated 1-bit feedback, long-term average power constraint $P$, $K$ users, temporal channel correlation coefficient $\rho$, threshold $\alpha$, and a prescribed transmission rate $R$, is given by* (30) *shown at the bottom of the page.*

*Proof:* Fixing the transmission rate $R$, the outage probability is (23) where

$$\epsilon_1 = \Pr[\log(1+v_\tau^2 P_1) < R \,|\, v^2 \geq \alpha],$$
$$\epsilon_0 = \Pr[\log(1+v_\tau^2 P_0) < R \,|\, v^2 < \alpha],$$

are the probabilities of outage conditioned on the feedback bits. Similar to the proof of Proposition 3, and using the conditional pdf (6), and (2.20), (B.19) in [30], it is straightforward to show,

$$\epsilon_1 = Q_1\left(\sqrt{\frac{\mu}{P_1}}, |\rho|\sqrt{\nu}\right) - e^{\alpha - \frac{e^R-1}{P_1}} Q_1\left(\sqrt{\frac{\mu}{P_1}}|\rho|, \sqrt{\nu}\right),$$

and

$$\epsilon_0 = \frac{1}{1-e^{-\alpha}}\left(1 - e^{-\frac{e^R-1}{P_0}} - e^{-\alpha} Q_1\left(\sqrt{\frac{\mu}{P_0}}, |\rho|\sqrt{\nu}\right)\right.$$
$$\left. + e^{-\frac{e^R-1}{P_0}} Q_1\left(\sqrt{\frac{\mu}{P_0}}|\rho|, \sqrt{\nu}\right)\right), \tag{31}$$

$$\epsilon = \left(1-(1-e^{-\alpha})^K\right)\left[Q_1\left(\sqrt{\frac{\mu}{P_1}}, |\rho|\sqrt{\nu}\right) - e^{\alpha - \frac{e^R-1}{P_1}} Q_1\left(\sqrt{\frac{\mu}{P_1}}|\rho|, \sqrt{\nu}\right)\right]$$
$$+ (1-e^{-\alpha})^{K-1}\left[1 - e^{-\frac{e^R-1}{P_0}} - e^{-\alpha} Q_1\left(\sqrt{\frac{\mu}{P_0}}, |\rho|\sqrt{\nu}\right) + e^{-\frac{e^R-1}{P_0}} Q_1\left(\sqrt{\frac{\mu}{P_0}}|\rho|, \sqrt{\nu}\right)\right], \tag{30}$$

$$\text{where } \mu = \frac{2(e^R-1)}{1-\rho^2} \text{ and } \nu = \frac{2\alpha}{1-\rho^2}.$$

where $\mu = \frac{2(e^R-1)}{1-\rho^2}$ and $\nu = \frac{2\alpha}{1-\rho^2}$, so that the probability of outage $\epsilon$ in (30) under the long-term average power constraint easily follows. ∎

*B.2. Diversity-Multiplexing Tradeoff*

**Proposition 6** *The DMT of the considered transmission scheme with outdated 1-bit feedback, given $K$ users, channel correlation coefficient $|\rho| < 1$, and long-term average power constraint, is given by*

$$d_o(r) = (1-r)^+. \tag{32}$$

*Proof:* In (30), the outage probability of the 1-bit feedback scheme with outdated CSI is the summation of two terms. The first term is the outage probability for the case that at least one user has a channel gain above the threshold when measure the channel; the second term corresponds to the outage probability for the case when all users are with channel gains below the threshold. At the asymptotic high SNR regime, the case that has a worse DMT (slower decreasing outage probability) determines the scheme's overall DMT. A detailed and rigorous proof is shown in Appendix D.

***Remark 4:*** Our result complements and reinforces the result in [22], and shows that even when power allocation is considered, it can not boost the DMT of the 1-bit feedback scheme with delay. Intuitively, the 1-bit feedback is used to indicate the base station the users' channel conditions (above or lower than the threshold), and with the uncertainty brought by the feedback delay, the outdated 1-bit feedback can not fulfill it. Thus the 1-bit feedback becomes meaningless when deriving the DMT, and the same DMT as of the SISO no feedback case is achieved for the outdated 1-bit feedback case. Note that although the 1-bit feedback scheme with delay does not help the DMT compared to the case of no CSI at the transmitter, it does improve the actual outage probability, especially at finite SNR regime. This is discussed in Sec. IV-C via numerical results.

*C. Numerical Results*

The outage probability $\epsilon$ of the broadcast channel with instantaneous 1-bit feedback is shown in Fig. 3 as a function of SNR $P$, for long-term average power constraint, different number of users $K$, and fixed outage rate $R = 3$ bits/sec/Hz. The cases for short-term average power constraint and different number of users are also shown for reference. For all cases, we choose the threshold $\alpha$ to guarantee that no outage occurs when there is at least one user with the channel gain above the threshold. As in the proof of Proposition 4, for the long-term average power constraint case, we choose $P_1 = P/2$, when there is a user with the channel gain above the threshold $\alpha$, and $P_0 = P_1/\Pr(N=0) = \frac{P}{2(1-e^{-\alpha})^K}$, when no user has a channel gain above the threshold. It is seen that in the high SNR regime, the slope of the outage probability curves increases with the number of users, indicating an increase in the diversity gain. Curves representing long-term power constraint (two-level power allocation) exhibit higher diversity gains than the corresponding curves for short-term power constraint. Note that in Fig. 3, the short-term average power constraint strategy achieves better outage performance at low

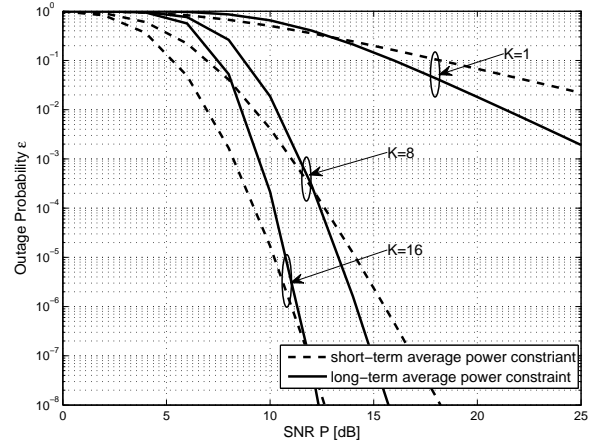

Fig. 3. Outage probability $\epsilon$ versus SNR $P$, for broadcast channels with instantaneous 1-bit feedback, both short-term and long-term average power constraints, different number of users ($K = 1, 8, 16$) and outage rate $R = 3$ bits/sec/Hz.

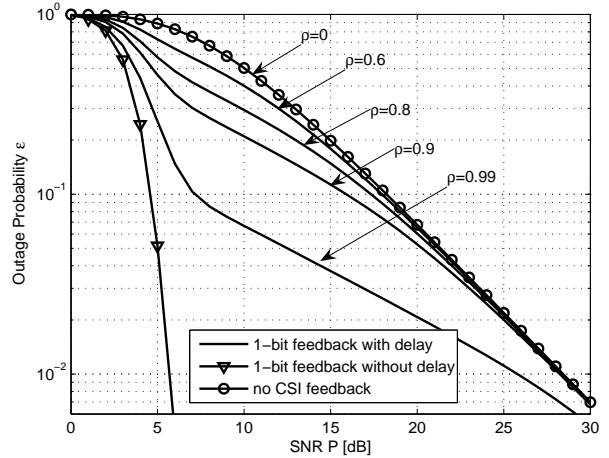

Fig. 4. Outage probability $\epsilon$ versus SNR $P$ for broadcast channels with outdated 1-bit feedback, different temporal channel correlation coefficients $\rho$, 16 users and outage rate $R = 3$ bits/sec/Hz. Outage probability with instantaneous 1-bit CSI feedback, and no CSI ($\rho = 0$) are also shown for reference.

SNR regime compared to long-term average power constraint case. This is because of the specific power allocation scheme that we use in Proposition 4, and we do not fully utilize the total long-term average transmission power $P$, especially in low SNR regime (see (36)).

Fig. 4 shows the outage probability $\epsilon$ of the outdated 1-bit feedback scheme as a function of SNR $P$, for different temporal channel correlation coefficients $\rho$, finite number of users $K = 16$ and fixed rates $R = 3$ bits/sec/Hz. The curves for the outage probability of the 1-bit feedback scheme without delay, for no CSI feedback, are also shown for reference. It can be seen that in the presence of delay, the 1-bit feedback scheme shows the same diversity order as of the transmission scheme with no CSI at the transmitter. However, the knowledge of partial CSI through the 1-bit does benefit the outage probability of the scheme for finite SNR. In fact, increasing the channel correlation coefficient $\rho$, we obtain better outage performance for the 1-bit feedback scheme with delay.

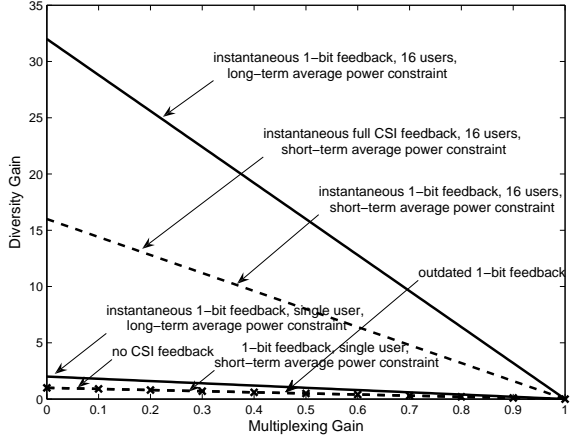

Fig. 5. DMT of broadcast channels with 16 users, for different feedback and average power constraint schemes. The tradeoff for point-to-point communication with 1-bit feedback is also shown for reference.

DMT curves for different feedback schemes, and short- and long-term power constraints are shown in Fig. 5. The curves are generated as follows: 1-bit feedback with long-term average power constraint (29), 1-bit feedback with short-term average power constraint [22], full CSI feedback with short-term power constraint [10], outdated 1-bit feedback (32), and no CSI feedback [11]. The DMT for point-to-point communication with 1-bit feedback [26] is also shown for reference. As can be seen, with 16 users and instantaneous 1-bit feedback in the system, a large improvement in the diversity gain can be observed compared to the single user 1-bit feedback case. Besides, diversity gains of 16 and 32 are obtained for short-term and long-term average power constraints, respectively, showing that power allocation is the key factor to double the diversity gain. On the other hand, the curve for the outdated 1-bit feedback shows that no advantage in terms of DMT is achieved compared to the no CSI feedback case when feedback delay is considered, though actual improvement in terms of outage probability is available, as shown in Fig. 4.

## V. CONCLUDING REMARKS

In this work we investigate the ergodic sum-rate and outage probability of a single-antenna broadcast channel with 1-bit instantaneous or outdated feedbacks. First, for the ergodic case, a closed-form expression of the achievable ergodic sum-rate that holds for any number of users, temporal channel correlation coefficient, and threshold, has been derived, along with simple upper and lower bounds. It is shown that reducing the CSI feedback to 1 bit, and even when subject to feedback delay, the scaling law of the ergodic sum-rate is the same as that of a system with full CSI at the transmitter. However, the feedback delay entails an ergodic sum-rate degradation that is quantified for both cases of large number of users and low SNR. For the outage case, a long-term average power constraint is assumed for the 1-bit feedback scheme, where different power is adaptively allocated to the transmission according to the CSI feedback from all users. We derive the expression for the outage probability, and evaluate the system's asymptotic behavior by finding the DMT. It is shown that by employing two power levels, the DMT is doubled for the instantaneous 1-bit feedback case, and an improvement in terms of outage probability is observed for the outdated feedback case. Our results are not only analytically and numerically meaningful, but also shed light on the system design aspects of simplified limited feedback link adaptation and Adaptive Modulation and Coding (AMC) schemes such as in HSDPA and WiMAX.

## APPENDIX

### A. Derivation of (8):

Starting from (2), since the integrand is positive, we replace the lower limit of integration with $\sqrt{\alpha}$, obtaining the following lower bound

$$R(\alpha, \rho, P, K) \geq \left(1-(1-e^{-\alpha})^K\right) \int_{\sqrt{\alpha}}^{\infty} \log(1+z^2 P) 2z \\ \times e^{-z^2+\alpha} Q_1\left(\frac{\sqrt{2}\rho}{\sqrt{1-\rho^2}}z, \frac{\sqrt{2\alpha}}{\sqrt{1-\rho^2}}\right) dz. \quad (33)$$

Then, the integration variable $z$ in the increasing function of log is replaced by the lower limit of the integration in (33), yielding the strict lower bound

$$\begin{aligned} R(\alpha, \rho, P, K) &> \left(1-(1-e^{-\alpha})^K\right) \log(1+\alpha P) \int_{\sqrt{\alpha}}^{\infty} 2z e^{-z^2+\alpha} \\ &\quad \times Q_1\left(\frac{\sqrt{2}\rho}{\sqrt{1-\rho^2}}z, \frac{\sqrt{2\alpha}}{\sqrt{1-\rho^2}}\right) dz \\ &= \left(1-(1-e^{-\alpha})^K\right) \log(1+\alpha P) \\ &\quad \left\{1 + Q_1\left(\frac{\rho\sqrt{2\alpha}}{\sqrt{1-\rho^2}}, \frac{\sqrt{2\alpha}}{\sqrt{1-\rho^2}}\right) \right. \\ &\quad \left. - Q_1\left(\frac{\sqrt{2\alpha}}{\sqrt{1-\rho^2}}, \frac{\rho\sqrt{2\alpha}}{\sqrt{1-\rho^2}}\right)\right\} \\ &= R_{\text{low}}(\alpha, \rho, P). \quad (34) \end{aligned}$$

The first equality in (34) follows from Eq.(B.18) of [30].

### B. Proof of (11):

As $B \to \infty$, using the asymptotic form of the zero-order modified Bessel function of the first kind, $Q_1(A, B)$ can be approximated as [36] (Eq.(A-27)),

$$\begin{aligned} Q_1(A, B) &\cong \int_B^{\infty} x \exp\left(-\frac{x^2+A^2}{2}\right) \frac{\exp(Ax)}{\sqrt{2\pi A x}} dx \\ &\cong \sqrt{\frac{B}{A}} \frac{1}{\sqrt{2\pi}} \int_B^{\infty} \exp\left(-\frac{(x-A)^2}{2}\right) dx \\ &= \sqrt{\frac{B}{A}} \Phi(B-A) \\ &\cong (2\pi AB)^{-1/2} \exp\left(-\frac{(B-A)^2}{2}\right), \quad (35) \end{aligned}$$

where $\Phi(t) \equiv \int_{-\infty}^{t} dx (2\pi)^{-1/2} \exp(-x^2/2)$.

Therefore, plugging (35) in (11), we easily obtain the limit we set out to prove.

### C. Proof of Proposition 4:

We look at the specific power allocation scheme, where we transmit $P_1 = P/2$, when there is at least one user above the threshold $\alpha$, and $P_0 = P_1/\Pr(N=0) = \frac{P}{2(1-e^{-\alpha})^K}$ when no user has a channel gain above the threshold. The average long-term power is then

$$\begin{aligned} P_{av} &= \Pr(N>0)P_1 + \Pr(N=0)P_0 \\ &= \left(1 - (1-e^{-\alpha})^K\right)\frac{P}{2} + \frac{P}{2} \leq P. \end{aligned} \quad (36)$$

We choose the threshold $\alpha$ to guarantee that there is no outage occur during the transmission when there is at least a user above the threshold $\alpha$, so that

$$R = \log(1 + \frac{P}{2}\alpha).$$

Thus the threshold is a function of power $P$,

$$\alpha = \frac{2(e^R - 1)}{P}. \quad (37)$$

With the chosen threshold $\alpha$, the probabilities of outage conditioned on the feedback bits are

$$\epsilon_1 = 0,$$

and

$$\epsilon_0 = \frac{1 - \exp(-\frac{e^R - 1}{P_0})}{1 - e^{-\alpha}} = \frac{1 - \exp\left(-\frac{2(e^R-1)(1-e^{-\alpha})^K}{P}\right)}{1 - e^{-\alpha}}.$$

Therefore the outage probability is

$$\begin{aligned} \epsilon &= \epsilon_1 \times \Pr(N>0) + \epsilon_0 \times \Pr(N=0) \\ &= \left(1-e^{-\frac{2(e^R-1)}{P}}\right)^{K-1}\left(1-e^{-\frac{2(e^R-1)\left(1-e^{-\frac{2(e^R-1)}{P}}\right)^K}{P}}\right). \end{aligned} \quad (38)$$

Setting the rate to $R = r \log P$, for multiplexing gain $r < 1$, $d(r)$ is

$$\begin{aligned} d(r) &= -\lim_{P \to \infty} \frac{\log \epsilon(\alpha, P, K, r\log P)}{\log P} \\ &= -\lim_{P \to \infty} \frac{\log \left(1 - e^{-\frac{2(e^{r\log P}-1)}{P}}\right)^{K-1}}{\log P} \\ &\quad - \lim_{P \to \infty} \frac{\log \frac{2(e^{r\log P}-1)}{P}}{\log P} - \lim_{P \to \infty} \frac{\log \left(1-e^{-\frac{2(e^{r\log P}-1)}{P}}\right)^K}{\log P} \\ &= 2K(1-r). \end{aligned} \quad (39)$$

Alternatively, if the multiplexing gain is $r \geq 1$, plug in $R = r \log P$ into (38), it is easily shown that the diversity is $d = 0$, thus completing the proof. ∎

### D. Proof of Proposition 6:

We first examine the first term in (30), which corresponds to the case of at least one user has the channel gain above the threshold when measure the channel. Under the same long-term average power constraint and the threshold that are used in Proposition 4, when $|\rho| < 1$, the diversity $d_{o1}$ is

$$d_{o1}(r) = -\lim_{P \to \infty} \frac{\log\{Q_1(a, |\rho|a) - Q_1(|\rho|a, a)\}}{\log P}, \quad (40)$$

where $a = \frac{\sqrt{2(e^R-1)}}{\sqrt{P_1(1-\rho^2)}}$.

When $0 \leq A < B$, $Q_1(A,B)$ can be bounded as [30] (Eq.(C.23)),

$$\exp\left(-\frac{(B+A)^2}{2}\right) \leq Q_1(A,B) \leq \exp\left(-\frac{(B-A)^2}{2}\right), \quad (41)$$

and

$$Q_1(A,B) \geq 1 - \frac{1}{2}\left[\exp\left(-\frac{(B-A)^2}{2}\right) - \exp\left(-\frac{(B+A)^2}{2}\right)\right], \quad (42)$$

when $0 \leq B < A$ [30] (Eq.(C.24)). Plugging (41) and (42) in (40), we obtain

$$d_{o1}(r) \leq -\lim_{P \to \infty} \frac{\log\left(1 - \frac{3}{2}e^{-\frac{a^2(1-|\rho|^2)}{2}} + \frac{1}{2}e^{-\frac{a^2(1+|\rho|^2)}{2}}\right)}{\log P}. \quad (43)$$

With the Taylor Series expansion of the exponential functions in the above equation, omitting the higher orders (which does not affect the DMT), and plug in $a = \frac{\sqrt{2(e^R-1)}}{\sqrt{P_1(1-\rho^2)}}$, it easily follows that

$$d_{o1}(r) \leq (1-r)^+, \quad \text{when } r < 1. \quad (44)$$

Since a SISO case at least achieves a DMT of $d = (1-r)^+$ even without any CSI at the transmitter, we conclude that $d_{o1} = (1-r)^+$.

We then check the second term in (30). By choosing the same threshold $\alpha$, a diversity gain of $K-1$ is at least achieved from the term $(1-e^{-\alpha})^{K-1}$. Thus the scheme's overall DMT is determined by the DMT of the case when at least one user has the channel gain above the threshold when measure the channel, and $d_o(r) = d_{o1}(r) = (1-r)^+$. ∎

### REFERENCES

[1] E. Biglieri, J. Proakis, and S. Shamai, "Fading channels: Information-theoretic and communications aspects," *IEEE Trans. Inform. Theory*, vol. 44, no. 6, pp. 2619–2692, Oct. 1998.

[2] L. Li and A. J. Goldsmith, "Capacity and optimal resource allocation for fading broadcast channels, - part I: Ergodic capacity," *IEEE Trans. Inform. Theory*, vol. 47, no. 3, pp. 1083–1102, Mar. 2001.

[3] ——, "Capacity and optimal resource allocation for fading broadcast channels, - part II: Outage capacity," *IEEE Trans. Inform. Theory*, vol. 47, no. 3, pp. 1103–1127, Mar. 2001.

[4] D. Tuninetti and S. Shamai, "The capacity region of two user fading broadcast channels with perfect channel state information at the receivers," in *Proc. IEEE Int. Symp. Inf. Theory*, Yokohama, Japan, June 2003, p. 345.

[5] W. Zhang, S. Kotagiri, and J. N. Laneman, "On downlink transmission without transmit channel state information and with outage constraints," *IEEE Trans. Inform. Theory*, to appear.

[6] R. Knopp and P. Humblet, "Information capacity and power control in single cell multiuser communications," in *Proc. IEEE Int. Conf. Commun.*, Seattle, WA, June 1995, pp. 331–335.

[7] D. N. C. Tse, "Optimal power allocation over parallel Gaussian broadcast channels," in *Proc. IEEE Int. Symp. Inf. Theory*, Ulm, Germany, June 1997, p. 27.

[8] P. Viswanath, D. N. C. Tse, and R. Laroia, "Opportunistic beamforming using dumb antennas," *IEEE Trans. Inform. Theory*, vol. 48, no. 6, pp. 1277–1294, June 2002.

[9] K. Josiam, D. Rajan, and M. D. Srinath, "Diversity multiplexing tradeoff in multiple antenna multiple access channels with partial CSIT," in *Proc. IEEE Global Telecommun. Conf. (GLOBECOM)*, Washington, DC, Nov 2007, pp. 3210–3214.

[10] K. Josiam and D. Rajan, "Multiuser diversity in wireless networks: A diversity-multiplexing perspective," in *Proc. Conf. on Inf. Sci. and Sys.*, Johns Hopkins University, Maryland, Mar. 2007, p. 419.

[11] L. Zheng and D. N. C. Tse, "Diversity and multiplexing: A fundamental tradeoff in multiple antenna channels," *IEEE Trans. Inform. Theory*, vol. 49, no. 5, pp. 1073–1096, May 2003.

[12] D. Gesbert and S. Alouini, "How much feedback is multi-user diversity really worth," in *Proc. IEEE Int. Conf. Commun.*, Paris, France, June 2004, pp. 234–238.

[13] V. Hassel, M.-S. Alouini, G. Øien, and D. Gesbert, "Exploiting multiuser diversity using multiple feedback thresholds," in *Proc. IEEE Veh. Technol. Conf.*, Stockholm, Sweden, May 2005, pp. 1302–1306.

[14] M. Johansson, "Benefits of multiuser diversity with limited feedback," in *Proc. IEEE Int. Workshop Signal Proc. Adv. in Wireless Commun. (SPAWC)*, Rome, Italy, June 2003, pp. 155–159.

[15] F. Florén, O. Edfors, and B.-A. Molin, "The effect of feedback quantization on the throughput of a multiuser diversity scheme," in *Proc. IEEE Global Telecommun. Conf. (GLOBECOM)*, San Francisco, CA, Dec. 2003, pp. 497–501.

[16] Y. Yu and G. B. Giannakis, "Opportunistic medium access for wireless networking adapted to decentralized CSI," *IEEE Trans. Wireless Commun.*, vol. 5, no. 6, pp. 1445–1455, June 2006.

[17] S. Y. Park, D. Park, and D. Love, "On scheduling for multiple-antenna wireless networks using contention-based feedback," *IEEE Trans. Commun.*, vol. 55, no. 6, pp. 1174–1190, June 2007.

[18] D. Love, R. W. Heath, V. K. N. Lau, D. Gesbert, B. D. Rao, and M. Andrews, "An overview of limited feedback in wireless communication systems," *IEEE J. Select. Areas Commun.*, vol. 26, no. 8, pp. 1341–1365, Oct. 2008.

[19] S. Sanayei and A. Nosratinia, "Exploiting multiuser diversity with only 1-bit feedback," in *Proc. IEEE Wireless Commun. and Networking Conf. (WCNC)*, New Orleans, LA, Mar. 2005, pp. 978–983.

[20] O. Somekh, A. M. Haimovich, and Y. Bar-Ness, "Sum-rate analysis of downlink channels with 1-bit feedback," *IEEE Commun. Lett.*, vol. 11, no. 2, pp. 137–139, Feb. 2007.

[21] J. Diaz, O. Simeone, and Y. Bar-Ness, "Asymptotic analysis of reduced-feedback strategies for MIMO Gaussian broadcast channels," *IEEE Trans. Inform. Theory*, vol. 54, no. 3, pp. 1308–1316, Mar. 2008.

[22] Y. Nam, J. Zhang, H. El-Gamal, and T. Reid, "Opportunistic communications with distorted CSIT," in *Int. Symp. on Wireless Commun. Systems (ISWCS)*, Valencia, Spain, Sept. 2006, pp. 16–20.

[23] Q. Ma and C. Tepedelenlioğlu, "Practical multiuser diversity with outdated channel feedback," *IEEE Trans. Veh. Technol.*, vol. 54, no. 4, pp. 1334–1345, July 2005.

[24] V. Hassel, M.-S. Alouini, G. Øien, and D. Gesbert, "Rate optimal multiuser scheduling with reduced feedback load and analysis of delay effects," *EURASIP Journal on Wireless Commun. and Networking, special issue radio resource management in 3G+ systems*, Article ID 36424, 7 pages, 2006.

[25] B. Niu, O. Simeone, O. Somekh, and A. M. Haimovich, "On the sum-rate of broadcast channels with outdated 1-bit feedback," in *Proc. Asilomar Conf. Signals, Systems and Computers*, Pacific Grove, CA, Oct. 2006, pp. 201–205.

[26] T. T. Kim and M. Skoglund, "Diversity-multiplexing tradeoff in MIMO channels with partial CSIT," *IEEE Trans. Inform. Theory*, vol. 53, no. 8, pp. 2743–2759, Aug. 2007.

[27] A. Khoshnevis and A. Sabharwal, "On the asymptotic performance of multiple antenna channels with quantized feedback," *IEEE Trans. Wireless Commun.*, vol. 7, no. 10, pp. 3869–3877, Oct. 2008.

[28] Y. Peng, K. Josiam, and D. Rajan, "Diversity multiplexing tradeoff in MIMO frequency selective channels with partial CSIT," *IEEE Commun. Lett.*, vol. 12, no. 6, pp. 408–410, June 2008.

[29] W. C. Jakes, *Microwave Mobile Communications*. New York: Wiley, 1974.

[30] M. K. Simon, *Probability Distributions Involving Gaussian Random Variables: A Handbook for Engineers and Scientists*. Kluwer Academic Press, 2002.

[31] S. Verdú, "Spectral efficiency in the wideband regime," *IEEE Trans. Inform. Theory*, vol. 48, no. 6, pp. 1319–1343, June 2002.

[32] D. N. C. Tse and P. Viswanath, *Fundamentals of Wireless Communications*. Cambridge, U.K.: Cambridge Univ. Press, 2005.

[33] L. Li, N. Jindal, and A. J. Goldsmith, "Outage capacities and optimal power allocation for fading multiple-access channels," *IEEE Trans. Inform. Theory*, vol. 51, no. 4, pp. 1326–1347, Apr. 2005.

[34] J. Luo, R. Yates, and P. Spasojević, "Service outage based power and rate allocation for parallel fading channels," *IEEE Trans. Inform. Theory*, vol. 51, no. 7, pp. 2594–2611, July 2005.

[35] B. Niu and A. M. Haimovich, "On the outage and diversity-multiplexing tradeoff of broadcast channels with 1-bit feedback," in *Proc. IEEE Int. Conf. Acoust., Speech, Signal Process.*, Dallas, TX, Mar. 2010, submitted.

[36] A. H. Nuttall, "Some integrals involving the Q-function," Technical Report TR4297, Naval Underwater System Center, Tech. Rep., Apr. 1972.